\documentclass[final,11pt,3p]{elsarticle}
 \usepackage{amsmath}
\usepackage{color}
\usepackage{hyperref}
\usepackage{cleveref}
\usepackage{graphicx}
\usepackage{amssymb}
\usepackage{amsthm}
\newtheorem{theorem}{Theorem}

\newtheorem{definition}{Definition}

\usepackage{lineno}
\usepackage{makecell}

\usepackage[ruled,noend,noline,linesnumbered,noresetcount]{algorithm2e}
\usepackage{booktabs}
\let\oldnl\nl
\newcommand{\nonl}{\renewcommand{\nl}{\let\nl\oldnl}}

\SetKwInput{Initial}{Initial}
\SetKwInput{Process}{Process 1}
\SetKwFor{Upon}{upon}{}{}
\SetKwFor{Wait}{wait}{}{}
\makeatletter
\def\ps@pprintTitle{%
 \let\@oddhead\@empty
 \let\@evenhead\@empty
 \def\@oddfoot{}%
 \let\@evenfoot\@oddfoot}
\makeatother
\begin{document}
\begin{frontmatter}
\title{A Practical Semi-Quantum Signature Protocol with  Improved Eavesdropping Detection}
\pdfstringdefDisableCommands{%
  \def\corref#1{}%
  \def\cnotenum{}%
}

\author[label1]{ZENGYU PANG}

\author[label1]{HUA XIANG\corref{cor1}}
\ead{hxiang@whu.edu.cn}

\address[label1]{School of Mathematics and Statistics, Wuhan University, Wuhan, 430072, China}

\cortext[cor1]{Corresponding author}

\begin{abstract}
Semi-quantum signature (SQS) schemes aim to enable quantum signature functionality in scenarios where only a subset of participants possess full quantum capabilities, thereby improving practical deployability while preserving quantum security advantages. Within this framework, we present a practical SQS protocol based on Bell states. The protocol is designed so that only the signer requires full quantum capability, significantly alleviating the quantum burden on the remaining participants. To strengthen security in semi-quantum environments, we incorporate an improved eavesdropping-detection mechanism that more effectively detects tampering. Compared with many existing schemes, which do not explicitly consider tampering of already generated signatures in their unforgeability analyses, the proposed protocol is designed to remain secure in the presence of such tampering.
\end{abstract}


\end{frontmatter}

\section{Introduction}\label{sec:introduction}
The rapid development of quantum technologies has brought both fundamental challenges and new opportunities to modern cryptography. On the one hand, quantum parallelism and quantum algorithms can threaten the computational assumptions that underlie many classical public-key schemes, thereby creating an urgent demand for cryptographic primitives with stronger security guarantees \cite{shor1999polynomial,gennaro2000rsa,johnson2001elliptic,huang2020quantum,grover1997quantum}. On the other hand, inherent properties of quantum mechanics, including the no-cloning theorem, measurement-induced state collapse, and quantum entanglement, provide entirely new tools for constructing communication and authentication mechanisms with information-theoretic security \cite{bennett2014quantum,bennett1984update,bennett1992quantum}. Following the introduction of quantum key distribution (QKD), researchers began to explore the direct use of quantum resources to realize signature and authentication functionalities, which led to the emergence of quantum digital signature (QDS) schemes \cite{gottesman2001quantum,zeng2002arbitrated}.

Quantum digital signature (QDS) is a cryptographic primitive built on the fundamental principles of quantum mechanics. It aims to provide security properties such as unforgeability and undeniability for message authentication. Unlike classical digital signature schemes that rely on computational hardness assumptions, QDS exploits intrinsic quantum features, including the no-cloning theorem and measurement disturbance, so that any unauthorized copying or tampering with the signature states inevitably introduces detectable discrepancies. As a result, QDS protocols can in principle offer information-theoretic security under appropriate models \cite{yang2011arbitrated,luo2012quantum,zou2013security,su2013improved,xin2019efficient,xin2019security,zheng2020arbitration,ding2022security}.

However, most existing QDS schemes assume that all participants possess full quantum capabilities, including the ability to prepare arbitrary quantum states, perform measurements in multiple bases, and store quantum information. Such requirements significantly increase implementation complexity and cost, thereby limiting practical deployment. To bridge the gap between theoretical constructions and realistic scenarios, the semi-quantum paradigm was introduced. Initially developed in the context of semi-quantum key distribution (SQKD), this idea has subsequently been extended to the design of semi-quantum signature (SQS) schemes \cite{boyer2007quantum,boyer2017experimentally,zou2009semiquantum}. In this framework, only certain participants need full quantum capabilities, while the remaining parties are restricted to limited  quantum operations—such as preparing and measuring qubits in the Z basis, reflecting incoming particles, or reordering them. Because the classical participant does not require advanced quantum devices (for example, universal state generators, quantum memory, or multi-basis measurement apparatus), the hardware requirements are substantially reduced. In practice, the classical party only needs a reflector, a delay/reordering device, and a simple detector capable of Z-basis measurements. Consequently, semi-quantum protocols reduce implementation complexity and cost, making them more practical when one party lacks full quantum capability.

Recently, many novel semi-quantum signature (SQS) protocols have been proposed \cite{zhao2019semi,zheng2020semi, chen2020offline,xia2022semi,zhao2023novel,yang2025semi,yang2022semi,yang2022cryptanalysis,he2023semi,zhang2023semi,zhang2024bell,shang2026semi}. In 2019, Zhao et al. proposed a semi-quantum bi-signature scheme based on W states for signing the same message, and ensured its security via W-state teleportation combined with a semi-quantum key distribution (SQKD) protocol \cite{zhao2019semi}. However, Yang et al. later pointed out that this method is vulnerable to forgery attacks \cite{yang2022semi}. In 2020, Zheng et al. introduced a semi-quantum proxy signature scheme that employs quantum-walk–based teleportation to generate entanglement during the signing process, thereby reducing the need for pre-prepared quantum resources and complicated operations \cite{zheng2020semi}. Also in 2020, Chen et al. proposed an offline arbitrated semi-quantum signature (ASQS) scheme based on four-particle cluster states. This scheme enables classical parties to sign messages with the assistance of a quantum arbitrator, simplifies transmission, and protects against forgery and repudiation \cite{chen2020offline}. In 2022, Xia et al. proposed an SQS scheme based on Einstein–Podolsky–Rosen (EPR) steering, which allows classical participants to sign and communicate securely while verifying identities and resisting attacks \cite{xia2022semi}. In 2023, He et al. proposed a semi-quantum ring-signature protocol based on multi-particle GHZ states, enabling users with limited quantum capability to sign messages securely and anonymously \cite{he2023semi}. In the same year, Zhang et al. proposed a designated-verifier SQS based on GHZ-like states \cite{zhang2023semi}. The protocol aims to reduce quantum resource consumption while preserving security and enabling dispute resolution through a semi-trusted third party. Subsequently, in 2024, they presented a Bell-state-based SQS scheme that is secure against quantum adversaries and more practical, since only the signer is required to possess quantum capabilities \cite{zhang2024bell}. More recently, Shang et al. proposed a semi-quantum multi-signature protocol in which all signers and the verifier are classical. The protocol uses only single qubits and simple permutation operations \cite{shang2026semi}.

Despite the progress achieved in recent semi-quantum signature schemes, there remains room for improvement in both practical implementability and  security.
In this paper, we propose a novel SQS protocol with the following main advantages:

(1) We improve the eavesdropping-detection strategy commonly adopted in existing semi-quantum protocols and incorporate it into our SQS framework. The enhanced detection mechanism strengthens the protocol’s ability to identify malicious interference during transmission and enables more effective resistance against tampering attacks.

(2) In our SQS protocol, only the signer is required to possess full quantum capabilities, while both the signature receiver and the trusted third party are classical participants. This setting significantly reduces the practical requirements for implementation and makes the protocol more feasible in realistic scenarios where fully quantum users are limited.

(3) The SQS protocol relies solely on Bell states as quantum resources. Compared with schemes that require multi-particle entangled states or complex entanglement structures, the use of Bell states simplifies state preparation and experimental realization, thereby improving overall practicality and reducing implementation cost.

(4) Many existing quantum signature schemes, when proving unforgeability, overlook tampering attacks on already generated valid signatures, which may lead to potential security vulnerabilities. In contrast, the proposed SQS protocol remains secure even in the presence of such tampering attacks.

The remainder of this paper is organized as follows. In Section 2, we present an improved eavesdropping-detection mechanism tailored for semi-quantum protocols. Section 3 formally introduces our proposed SQS protocol. In Section 4, we provide a comprehensive analysis of its security and efficiency. Section 5 compares our scheme with existing related protocols to highlight its advantages and distinctive features. Finally, Section 6 concludes the paper.

\section{Eavesdropping Detection in Semi-Quantum Scheme }

In this section, we discuss eavesdropping detection in the semi-quantum setting. We first briefly introduce the two types of participants involved in the semi-quantum framework in order to clarify the setting and assumptions adopted in our scheme.

In the semi-quantum model, participants are of two types: quantum parties  and classical parties. The quantum party has full quantum functionality, such as the ability to prepare arbitrary quantum states, perform general quantum operations, and measure qubits in different bases. In contrast, the classical party is restricted to a predefined set of operations: (1) preparing and sending qubits in the computational (Z) basis $\{|0\rangle, |1\rangle\}$; (2) measuring qubits only in the Z basis; (3) reflecting incoming qubits without disturbing their quantum states; and (4) temporarily delaying or reordering qubits.

During the transmission of quantum states, an eavesdropping-detection mechanism is typically introduced to determine whether there is interference from a third party. A common approach is for the sender  to randomly insert decoy states prepared in the Z and X bases into the quantum sequence. After the receiver confirms receipt of the sequence, the sender  announces the positions and content of the decoy states. The receiver then measures these decoy states and estimates the error rate to determine whether eavesdropping has occurred \cite{bennett2014quantum}.

However, when classical parties are involved, the above method cannot be applied, because a classical party cannot prepare decoy states or perform measurements in the X basis. Consequently, a common approach is for the receiver, Bob, to directly reflect the particles back to the sender, Alice, who then performs the corresponding measurements and checks the error rate  \cite{zhang2024bell,tian2021efficient,zhang2023semi}.

In fact, this method has potential security risks. As pointed out by He et al., the direct reflection approach is vulnerable to the DCNA attack \cite{he2024security}. Essentially, this is an entanglement-based measurement attack that allows an adversary to entangle the transmitted particles with their own ancilla, enabling them to extract useful information through later measurements. To address this issue, He et al. propose the following approach. After Bob receives the sequence containing the decoy states, he first measures the decoy states prepared in the Z basis according to the positions announced by Alice. He then extracts all the decoy states, reorders them, and sends them back to Alice. Upon receiving the returned decoy states, Alice performs measurements according to the reordering information later announced by Bob. By comparing the measurement results with the initially prepared decoy states, Alice estimates the error rate. The effectiveness of the above method in preventing entanglement attacks lies in the fact that, after the decoy particles are reordered, a third party can no longer determine the correspondence between the ancilla they introduced and the returned decoy particles. As a result, Alice can successfully detect the resulting errors.

Nevertheless, the above method still has a vulnerability: it cannot resist tampering attacks. To illustrate this, let us first revisit the original direct-reflection approach for decoy states. In addition to the DCNA attack pointed out by He et al., this method is also susceptible to tampering attacks as follows:
 \[\text{Alice}
\xrightarrow[\text{Eve}]{U\{|0\rangle,|1\rangle,|+\rangle,|-\rangle\}_{decoy}} \text{Bob}\xrightarrow[\text{Eve}]{U^{\dagger}U\{|0\rangle,|1\rangle,|+\rangle,|-\rangle\}_{decoy}}\text{Alice}.\]Here, an attacker can modify the particle sequence during Alice’s initial transmission and later correct these modifications when Bob reflects the decoy states back to Alice. As a result, Alice is unable to detect any errors.

In the method proposed by He et al., Bob first measures the received decoy states prepared in the Z basis and only then returns all the decoy states to Alice. At first glance, since the adversary does not know Bob’s measurement outcomes, it seems difficult to correct the previous tampering during the transmission of the decoy states back to Alice. The detailed procedure is as follows: \[\text{Alice}
\xrightarrow[\text{Eve}]{U\{|0\rangle,|1\rangle,|+\rangle,|-\rangle\}_{decoy}} \text{Bob}\xrightarrow[\text{Eve}]{U^{\dagger}\text{measured}(U\{|0\rangle,|1\rangle\}_{decoy})\,\text{and}\,U^{\dagger}U\{|+\rangle,|-\rangle\}_{decoy}}\text{Alice}.\]
Assume that during the first transmission Eve applies an operation $U$ to all quantum states, and during the second transmission she applies $U^{\dagger}$ again to reverse the tampering. For decoy states prepared in the X basis, this attack can indeed avoid being detected. However, for the decoy states in the Z basis, the measurement outcomes after the first tampering are unknown to Eve. As a result, she cannot faithfully restore the states to their original form, and the induced errors can be detected by Alice. However, there exist specific tampering operations that can still bypass the detection described above. For example, if Eve chooses the Pauli-$X$ operation as the tampering operation, then for the Z-basis decoy states, \[X^{\dagger}\text{measured}(X\{|0\rangle,|1\rangle\}_{decoy})=\{|0\rangle,|1\rangle\}_{decoy}.\] Consequently, Alice cannot detect any errors.
\subsection{Improved Scheme}
To address the above problem, we propose the following improved scheme.

\textbf{Step 1:} Alice prepares two types of decoy sequences. The first sequence, $D_Z$, consists of Z-basis decoy states, where each particle is randomly selected from $\{|0\rangle, |1\rangle\}$. The second sequence, $D_X$, consists of X-basis decoy states, where each particle is randomly selected from $\{|+\rangle, |-\rangle\}$. Alice mixes $D_Z$ and $D_X$ to form the decoy sequence $D$, and randomly inserts $D$ into the quantum sequence sent to Bob.
Then, she records the positions and content of the two types of decoys as $LOC_{Z} (LOC_{X}$).

\textbf{Step 2:} Alice sends the quantum sequence containing the decoy particles $D$ to Bob.

\textbf{Step 3:} After receiving all particles from Alice, Bob sends a classical message to Alice confirming the reception.

\textbf{Step 4:} After receiving Bob's confirmation message, Alice publishes $LOC_{Z}$, i.e., the positions and content of the Z-basis decoy particles.

\textbf{Step 5:} After receiving the $LOC_{Z}$ message published by Alice, Bob performs the following operations:

(1) According to $LOC_{Z}$, the particles in $D_Z$ are measured in the Z basis and the measurement results are compared with $LOC_{Z}$ to detect errors. If the error rate exceeds a preset threshold, the protocol is aborted.

(2) Remove the entire decoy sequence $D$ (both $D_Z$ and $D_X$) from the received quantum sequence. Strictly speaking, the particles in $D_Z$ are the states after being measured by Bob.

(3) Randomly shuffle the decoy sequence $D$ and send the reordered decoy particles back to Alice.

\textbf{Step 6:} After receiving all particles, Alice sends a classical message to Bob confirming the reception.

 \textbf{Step 7:} Bob publishes the original order mapping  of the returned decoy sequence.

\textbf{Step 8:} Based on the mapping relationship provided by Bob, Alice restores the order of decoy particles, then performs measurements and compares the measurement results with the contents of $LOC_Z$ and $LOC_X$. If the error rate exceeds a predetermined threshold, the protocol is aborted.

The key improvement is that error-rate verification is performed by both Alice and Bob. This prevents a single party from shouldering the entire detection task and increases robustness against tampering. In this refined design, Bob first performs an error check on the Z-basis decoy states immediately after reception, which enables him to detect potential tampering before the particles are returned. Subsequently, Alice conducts a second round of verification after the reordered decoy sequence is sent back.

 There remain two types of attacks that Bob cannot detect independently. 
The first type arises when Eve measures the decoy particles in the Z basis during the forward transmission. Such a measurement does not disturb decoy states originally prepared in the Z basis, and hence Bob’s local error check on the Z-basis decoys cannot reveal this attack. However, the same measurement inevitably disturbs decoy states prepared in the X basis. Since Bob does not measure the X-basis decoys at this stage, detection must rely on Alice’s subsequent verification of the returned X-basis decoy states.  Note that Eve might try to tamper with the decoys when they are sent back to Alice to remove the effects of her earlier measurement. However, since Eve does not know the original information of the X-basis decoys prior to measurement, such an attempt is infeasible.

The second type is the entanglement-measurement attack. In this attack, Eve introduces an auxiliary system and performs a joint unitary operation to entangle her ancillary particle with the transmitted particle. She then performs measurements on the ancilla at an appropriate time to obtain information indirectly without being immediately detected. This attack cannot be detected solely by Bob’s initial measurement. Its defense primarily relies on Bob’s random reordering of the decoy particles, which destroys the correspondence between Eve’s probe and the transmitted states. A complete security analysis of this countermeasure is provided by He et al. \cite{he2024security}. Since our improvement preserves the same reordering mechanism, their proof remains applicable and need not be repeated here.

In summary, the two-sided verification structure prevents an attacker from compensating for earlier manipulations during the return transmission. Consequently, malicious interference cannot be concealed, and the tampering behavior described earlier can be reliably detected.

\section{Our SQS Protocol}

\begin{figure}[t]
    \centering
    \includegraphics[width=1.0\linewidth]{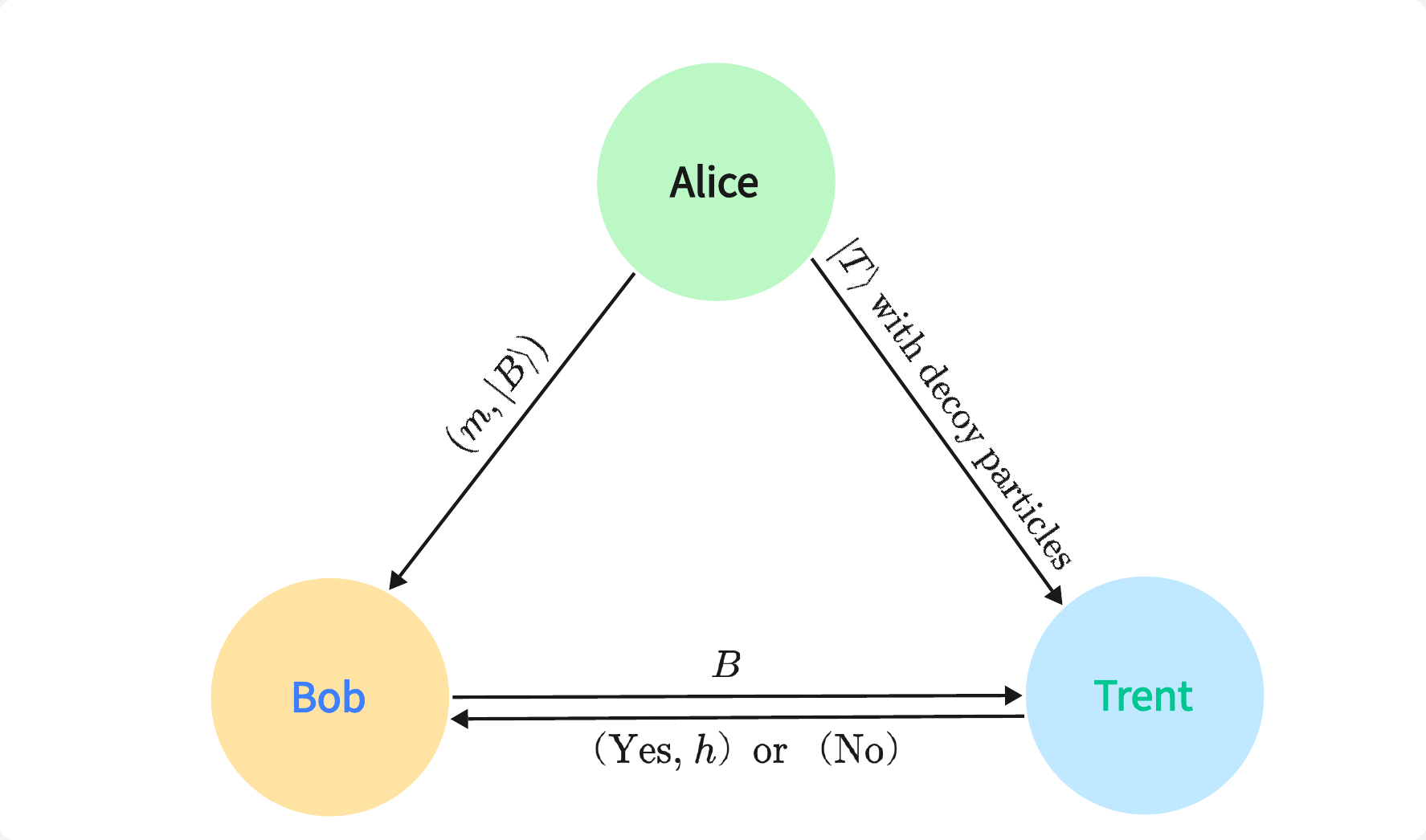}
    \caption{Signature process}
    \label{fig 4.1}
\end{figure}

Our SQS protocol involves three parties: Alice, Bob, and a trusted third party, Trent. Alice is the signer and is responsible for signing messages, while Bob is the recipient who verifies the signature with the assistance of the trusted third party Trent. Only Alice is a  quantum party with full quantum capabilities, Bob and Trent are classical parties. Let $m$ denote the message to be signed. The hash function is denoted as $h:\{0,1\}^* \rightarrow \{0,1\}^l$, which maps inputs of arbitrary length to fixed-length $l$-bit outputs. The symbol $\oplus$ denotes addition modulo $2$.

Figure \ref{fig 4.1} provides a schematic overview of the signing process. The detailed procedure is as follows.

\subsection{Initialization Phase}

In this phase, Alice and Trent share a private key through a semi-quantum QKD protocol, denoted as $K_{AT}=(k_1,k_2,\ldots,k_n)\in\{0,1\}^n$ \cite{boyer2007quantum,boyer2017experimentally,zou2009semiquantum}.

\subsection{Signing Phase}

In this phase, in order to sign the message $ m = (m_1, m_2, \ldots, m_n) \in \{0,1\}^n $, Alice performs the following operations.

\textbf{Step 1:} First, Alice performs a modulo-2 addition on $m$ and $K_{AT}$ to obtain $g$, which is denoted as \[g=m\oplus K_{AT}=(g_1, g_2, \ldots, g_n) \in \{0,1\}^n.\]

\textbf{Step 2}: Subsequently, the corresponding Bell state is prepared according to the value of $g_i\,(i\in n)$:

if $g_i=0$, prepare \[|\Phi^+\rangle=\frac{|0_T0_B\rangle+|1_T1_B\rangle}{\sqrt{2}};\]

if $g_i=1$, prepare \[|\Psi^+\rangle=\frac{|0_T1_B\rangle+|1_T0_B\rangle}{\sqrt{2}}.\]

We use $|T\rangle$ and $|B\rangle$ to denote the sequences formed by the first particles and the second particles, respectively.

\textbf{Step 3:} Alice sends the first-particle sequence $|T\rangle$ to Trent. To ensure a reliable transmission of the sequence, we adopt the eavesdropping-detection technique introduced earlier. In addition, we also intend to send $m$ to Trent. In practice, this  can be carried out using the eavesdropping-detection procedure.
Specifically, when preparing the Z-basis decoy states in $D_Z$, the bit information of $m$ can be embedded into them, for example by placing it at the beginning of the decoy sequence. Since Trent will measure these states to estimate the error rate, he can simultaneously  obtain the message $m$. 

Note that the method above does not provide confidentiality for $m$. Its intent is to make the decoy particles carry extra useful information, rather than serve solely as an eavesdropping-detection mechanism that is discarded after use.
For example, Eve could measure the quantum sequence $|T\rangle$ that Alice sends to Trent and forward the post-measurement sequence to Bob. After Bob confirms receipt, Alice would disclose the positions and content ($LOC_Z$) of the Z-basis decoys. Although the eavesdropping check might later reveal errors, Eve can already recover $m$ by combining the public $LOC_Z$ with her measurement results.
In our SQS setting $m$ is a public message, so this attack does not cause a problem. By contrast, some blind-signature schemes require that $m$ remain confidential, and therefore this approach would not be applicable in those cases \cite{zeng2002arbitrated,wen2009weak}.

Moreover, the eavesdropping-detection procedure can be further simplified. Owing to the existence of $K_{AT}$, Alice no longer needs to wait until Trent confirms receipt of the quantum sequence before announcing $LOC_Z$. Instead, she can encrypt $LOC_Z$ using a one-time pad with $K_{AT}$ and send the encrypted $LOC_Z$ together with the quantum sequence \cite{shannon1949communication,bellovin2011frank}. In this way, Steps 3 and 4 of the eavesdropping-detection process can be omitted. Similarly, when Trent returns the decoy particles to Alice, the same approach can be applied, allowing Steps 6 and 7 to be omitted as well. It is worth noting that, due to the one-time pad property, each encryption must use a distinct, non-reused key. In practice, this can be achieved by allocating a sufficiently long $K_{AT}$
 during the initialization phase and consuming it in disjoint segments for successive encryptions

\textbf{Step 4:} After receiving $|T\rangle$ with $m$, Trent first measures \(|T\rangle\) to obtain the measurement result \(T=(t_1, t_2, \ldots, t_n) \in \{0,1\}^n\), and then  computes \(g = K_{AT} \oplus m\) based on \(m\).

\textbf{Step 5:} Alice sends  $(m,|B\rangle)$  to Bob as a quantum signature.

\subsection{Verifying Phase}
In this phase, after Bob receives the signature from Alice, he verifies it with the help of a trusted third party, Trent. The specific steps are as follows:

\textbf{Step 1:} After receiving $(m,|B\rangle)$ from Alice, Bob measures $|B\rangle$ in the Z basis to obtain the result $B=(b_1, b_2, \ldots, b_n) \in \{0,1\}^n$, and then sends $B$ to Trent.

\textbf{Step 2:} After receiving \(B\), Trent performs the following verification using \(g\) : for any \(i \in n\), if \(g_i = 0\), he checks whether \(t_i\) is equal to \(b_i\); if they are equal, the verification passes, otherwise it fails. If \(g_i = 1\), the verification passes when \(t_i\) and \(b_i\) are different; otherwise, it fails (see Table \ref{tab:1}).

\textbf{Step 3:} Upon successful verification, Trent retains $(m, T, B)$ as evidence of the quantum signature, then computes the hash $h = h(m)$  and sends $(\mathrm{Yes},h)$ to Bob. If  verification fails, he sends (No) to Bob.

\textbf{Step 4}: After receiving $(\mathrm{Yes},h)$ from Trent, Bob checks whether $h = h(m)$. If they are equal, he accepts the signature.

\begin{table}[ht]
\centering
\renewcommand{\arraystretch}{1.5} 
\caption{Valid verification cases performed by Trent}
\label{tab:1}
\begin{tabular*}{0.5\linewidth}{@{\extracolsep{\fill}} p{1.5cm}|cccc}
\hline
 & Case 1 & Case 2 & Case 3 & Case 4 \\
\hline
\centering$g_i$ & 0 & 0 & 1 & 1 \\
\centering$t_i$ & 0 & 1 & 0 & 1 \\
\centering$b_i$ & 0 & 1 & 1 & 0 \\
\hline
\end{tabular*}
\end{table}

\section{Security  Analysis}
In this section, we provide a security analysis of the proposed SQS protocol.

\subsection{Correctness}
For a message $m=(m_1,\ldots,m_n)$, Alice first computes $g=m\oplus K_{AT}$ and encodes each bit $g_i$ into a Bell state, preparing $|\Phi^+\rangle=\frac{|00\rangle+|11\rangle}{\sqrt{2}}$ when $g_i=0$ and $|\Psi^+\rangle=\frac{|01\rangle+|10\rangle}{\sqrt{2}}$ when $g_i=1$. Alice sends the first-particle sequence $|T\rangle$ to Trent and the second-particle sequence $|B\rangle$ to Bob. In addition, she transmits the classical message $m$ to both Trent and Bob. In our construction, the transmission of $m$ to Trent is integrated into the eavesdropping-detection procedure: the bit information of $m$ is embedded into the Z-basis decoy states within the $|T\rangle$ sequence, so that Trent can recover $m$ when performing the corresponding measurements. Meanwhile, $m$ is sent directly to Bob as part of the signature tuple $(m, |B\rangle)$.

When Trent and Bob measure their respective particles in the Z basis, the intrinsic correlations of Bell states guarantee deterministic relationships between their outcomes: measuring $|\Phi^+\rangle$ yields identical bits while measuring $|\Psi^+\rangle$ yields opposite bits. This can also be understood as, for every position $i$, the measurement outcomes always satisfy $b_i=t_i\oplus g_i$ (see Table \ref{tab:1}). Since Trent can recompute $g$ using the shared key $K_{AT}$ and the received message $m$, his verification rule requires equality when $g_i=0$ and inequality when $g_i=1$, and this condition will be satisfied for all bits in the absence of interference.
 Upon successful verification, Trent computes the hash value $h = h(m)$ from the message $m$ he received from Alice and sends $(\mathrm{Yes}, h)$ to Bob. Bob then verifies the integrity of the received message by checking whether the hash value satisfies $h(m) = h$. If the verification succeeds, the signature is accepted; otherwise, it is rejected. Therefore, in the absence of interference, both Trent’s verification and Bob’s verification succeed, and the proposed SQS protocol satisfies the correctness property.

\subsection{Security of the private key}
In general, a quantum signature can be seen as the ciphertext produced by encrypting a message with the signer’s private key, and it could reveal some information about that key. For example, in our SQS framework the signature $|B\rangle$ of a message $m$ is formed by the second qubit of each Bell state, where the choice of Bell states is determined by $g = m \oplus K_{AT}$. By measuring and correlating the signer’s quantum signatures, an attacker may infer partial information about the private key $K_{AT}$. Therefore, a secure quantum signature scheme should provide information-theoretic security: the resulting quantum ciphertexts must be indistinguishable under chosen-plaintext attacks \cite{yang2012quantum,li2013quantum,menezes2018handbook}.

\begin{definition}\cite{yang2012quantum,li2013quantum}
 A quantum encryption scheme \(E\) is said to be information-theoretically secure (indistinguishable) if no distinguisher \(D\) can tell apart the ciphertexts of any two distinct plaintexts \(M\) and \(M'\) with a non-negligible advantage. Formally, for any distinguisher (D), any positive polynomial \(p(\cdot)\), and sufficiently large security parameter \(n\),
\[
\left|\Pr[D(E(M))=1]-\Pr[D(E(M'))=1]\right|<\frac{1}{p(n)}.
\]

\end{definition}
 This definition means that the ciphertexts produced by the scheme are indistinguishable under chosen-plaintext attacks, and thus reveal no useful information about the plaintext in the information-theoretic sense. 
 
 On the other hand, Li et al.  established a formal relation between information-theoretic security and the density operators of quantum ciphertexts. The precise statement is given in the theorem below \cite{li2013quantum}.

 \begin{theorem}
 Consider any two plaintexts $x$ and $y$, and denote the density operators of the corresponding ciphertext states $E(x)$ and $E(y)$ by $\rho_x$ and $\rho_y$. The encryption scheme achieves information-theoretic indistinguishability if, for every positive polynomial $p(\cdot)$ and all sufficiently large security parameters $n$, the ciphertext states remain negligibly close, namely
\[D(\rho_x,\rho_y) \le \frac{1}{p(n)}.\]
Here $D(\rho_x,\rho_y)$ represents the trace distance between the two quantum states, i.e. $D(\rho,\sigma)=\tfrac{1}{2}\mathrm{tr}\,|\rho-\sigma|,\,\text{where } |A|=\sqrt{A^\dagger A}.$

 \end{theorem}
  
Next, we show that both the signature sequence $|B\rangle$ and the decoy-assisted  sequence $|T\rangle$ satisfy the condition in Theorem 1, which implies the information-theoretic security of the private key $K_{AT}$.

We first recall a basic property of Bell states: the reduced density operator of either subsystem of a Bell state is the maximally mixed state. For example, for
$|\Phi^+\rangle=\frac{1}{\sqrt{2}}(|00\rangle+|11\rangle)$ and
$|\Psi^+\rangle=\frac{1}{\sqrt{2}}(|01\rangle+|10\rangle)$,
taking the partial trace over the second particle gives
\[\mathrm{tr}_B\left(|\Phi^+\rangle\langle\Phi^+|\right)=\frac{I_2}{2},\,
\mathrm{tr}_B\left(|\Psi^+\rangle\langle\Psi^+|\right)=\frac{I_2}{2}.\]
Therefore, regardless of which Bell state is prepared, the density operator of the particle sent to Bob is always the maximally mixed state $\frac{I_2}{2}$.
Then the density operator of the entire sequence $|B\rangle$ is a tensor product of maximally mixed states: \[\rho_{B}=\bigotimes_{i=1}^{n}\frac{I_2}{2}=\frac{I_{2^n}}{2^n},\]
where $n$ denotes the total number of qubits in $|B\rangle$.

We now consider the decoy-assisted sequence $|T\rangle$. For each decoy position, Alice randomly prepares one of the four states $\{|0\rangle,|1\rangle,|+\rangle,|-\rangle\}$.  Note that although we embed $m$ when constructing the Z-basis decoy states, the overall uniform distribution is not affected as long as the distribution of the remaining decoy states is properly adjusted. Due to the unconditional security of the one-time pad, an attacker cannot obtain the position and content information $LOC_Z$ \cite{shannon1949communication}. Hence, from the attacker’s perspective, the density operator of a single decoy qubit is
\[\frac{1}{4}\left(|0\rangle\langle0|+|1\rangle\langle1|+|+\rangle\langle+|+|-\rangle\langle-|\right)
=\frac{1}{4}(I_2+I_2)=\frac{I_2}{2}.\]
Like $|B\rangle$, the remaining qubits in $|T\rangle$ are  the reduced states $\frac{I_2}{2}$. Then, the entire sequence $|T\rangle$ is also a tensor product of maximally mixed states: \[\rho_{T}=\bigotimes_{i=1}^{n}\frac{I_2}{2}=\frac{I_{2^{n}}}{2^{n}}.\]

 For any two distinct messages $m$ and $m'$, the corresponding ciphertext states are identical for the attacker:
\[\rho^{m}_{B}=\rho^{m'}_{B}=\frac{I_{2^n}}{2^n},\,
\rho^{m}_{T}=\rho^{m'}_{T}=\frac{I_{2^{n}}}{2^{n}}.\]
Hence, the trace distance between the ciphertext states is
$D(\rho^m,\rho^{m'})=0,$
which clearly satisfies the condition $D(\rho_x,\rho_y)\le \frac{1}{p(n)}$. According to Theorem 1, the produced quantum ciphertexts are therefore information-theoretically indistinguishable. In fact, from a global perspective, the sequence $|T\rangle$, like $|B\rangle$, corresponds to a tensor product of maximally mixed states, and  no information about the private key can be inferred from it. The decoy particles inserted into $|T\rangle$ are completely independent of $K_{AT}$, and thus cannot reveal any useful information.

In addition to the quantum sequences $|B\rangle$ and $|T\rangle$, the classical values $m$ and $h$ are independent of the shared key $K_{AT}$. The string $B$ is merely the outcome of measuring $|B\rangle$ in the $Z$ basis; therefore, it also carries no information about $K_{AT}$. Moreover, when $K_{AT}$ is employed as a one-time pad (for example, to protect the decoy information), the unconditional security of the one-time pad guarantees that no information about $K_{AT}$ can be inferred from the ciphertext. Combining these observations, we can ensure the security of the private key.

\subsection{Unforgeability}
 We consider two cases. In the first case, suppose the attacker (Eve) attempts to forge a signature for a message \(m'\) from scratch. She  needs to construct a corresponding \(B'\) to be sent to Trent. Since Eve does not know \(K_{AT}\), she cannot compute \(g = K_{AT} \oplus m\), and thus cannot determine the Bell states chosen by Alice and Trent (Note that $g$ is computed using the message $m$ shared between Alice and Trent). Moreover, Eve does not know Trent’s measurement results on his entangled particles. Consequently, for each element $b'_i$ in $B'$, Eve can do no better than guess the correct value, and the probability that a single bit passes verification is $1/2$. Because the bits are verified independently, the probability that all $n$ elements simultaneously satisfy Trent’s verification condition is $(1/2)^n \to 0$ as $n \to \infty$. It should be noted that, due to the presence of the eavesdropping-detection mechanism, Eve cannot impersonate Alice to share the  particle sequence \(|T\rangle\) with Trent. Moreover, even if Eve were able to successfully forge \(|T\rangle\), without knowledge of \(K_{AT}\), which is required for Trent to compute \(g\), she would still be unable to predict the outcome of the verification process.

In the second case, suppose Eve attempts to tamper a legitimately generated signature. For the particle sequence $|T\rangle$ distributed from Alice to Trent, any interference during transmission will be revealed by the eavesdropping-detection mechanism. As a result, Eve cannot alter the values of $g$ or the corresponding measurement result $T$ that Trent uses.  For the particle sequence \(|B\rangle\) distributed by Alice to Bob, any modification would disturb the corresponding  entanglement relationship with $|T\rangle$. Since Trent’s verification explicitly checks the consistency between Bob’s measurement outcomes and his own results according to $g$, such disturbances will be detected with overwhelming probability. In addition, if Eve tampers with the classical message $m$ sent from Alice to Bob, Bob will detect the modification by computing the hash of his received message and comparing it with the value $h$ obtained from Trent. Any inconsistency between the two values indicates alteration and leads to rejection of the signature.

Combining the above two cases, we see that neither forgery from scratch nor tampering of a valid signature can succeed except with negligible probability. Therefore, the proposed protocol achieves unforgeability.

\subsection{Undeniability}

Undeniability means that the signer Alice cannot deny having sent a valid signature, and the receiver Bob cannot deny having received a valid signature.

For the former, based on the previous analysis of unforgeability, a signature forged by an attacker cannot pass Trent’s verification. Since Trent is trusted, once he successfully verifies the signature and stores $(m,T,B)$ as evidence, Alice cannot deny having sent a valid signature. Essentially, this is because only Alice and Trent possess $K_{AT}$.

For the latter, once Trent receives the message $B$ from Bob and the verification succeeds, it follows the unforgeability property that Bob must have obtained $B$ from the signature $(m,|B\rangle)$ sent by Alice.

Since the final step of the protocol requires Bob to perform hash verification, one might worry that Bob could tamper with $m$ while keeping the $B$ sent to Trent unchanged. In this case, if Bob ignores the hash $h$ from Trent and falsely claims that he received a valid signature, any later comparison with Trent’s stored evidence $(m,T,B)$ would reveal the inconsistency, allowing Alice to deny the tampered signature.

\subsection{Entanglement‑measurement attack}

The entanglement-measurement attack is a strategy in quantum cryptography in which an attacker introduces an auxiliary quantum system (a probe) and performs a joint unitary operation between the probe and the transmitted quantum state, thereby creating entanglement between them. Instead of directly measuring or replacing the state, the attacker keeps the probe and later measures it to extract information about the original quantum state, secret key, or encoded message. The purpose of this attack is to gain information indirectly through quantum correlations while attempting to minimize detectable disturbance.

In our SQS protocol, the attacker Eve may attempt to launch an entanglement-measurement attack on the quantum sequences \(|T\rangle\) and \(|B\rangle\). For \(|T\rangle\), due to the protection provided by our eavesdropping-detection mechanism, Eve cannot construct an effective probe without being detected. As for \(|B\rangle\), there is in fact no need to perform such an  attack. Since Bob measures \(|B\rangle\) immediately upon receipt, Eve could simply intercept and measure the sequence and then resend it to Bob. However, because our signature scheme achieves information-theoretic security, Eve cannot obtain any useful information from the \(|B\rangle\) sequence. Therefore, in summary, entanglement-measurement attacks are meaningless against our SQS protocol.

\subsection{Qubit Efficiency}

 To evaluate the communication cost of the proposed signature scheme we adopt the widely used qubit-efficiency metric \cite{cabello2000quantum}, defined as
 \[
 \eta=\frac{c}{q+b},
 \]
 where $c$ is the number of message bits that are successfully signed, $q$ is the total number of transmitted qubits required by the protocol, and $b$ is the number of auxiliary classical bits exchanged for verification. Qubits and classical bits transmitted solely for eavesdropping detection are excluded from the efficiency calculation.

 In the proposed SQS protocol, the size of the signed message $m$ is $n$ bits. The sequences $|T\rangle$ and $|B\rangle$ constitute the transmitted quantum states, totaling $2n$ qubits. The auxiliary classical bits include the measurement outcome string $B$ of length $n$ and the hash value $h$. Since the bit-length of $h$ is typically much smaller than $n$, the efficiency can be approximated as \[\eta \approx \frac{n}{2n+n} \approx 33\%.\]

\section{Comparisons }

This section presents a comparison between our protocol and other similar schemes. 

First, as an SQS protocol, our SQS protocol requires only the signer to possess quantum capabilities. By contrast, many protocols  require two quantum parties \cite{zhao2019semi,xia2022semi,zhang2023semi}. This means our protocol places lower demands on quantum hardware and is simpler to implement. Second, we have improved the existing eavesdropping-detection techniques for semi-quantum protocols, removing their vulnerability to tampering attacks. The older direct-reflection detection methods used in \cite{xia2022semi,zhang2023semi,zhang2024bell} are susceptible to such tampering. Moreover, our protocol uses only simple Bell states, whereas protocols like \cite{zhao2019semi,zhang2024bell,zhang2023semi} employ three-particle entangled states, \cite{chen2020offline} uses four-particle entangled states, and \cite{xia2022semi} uses Bell and other two-particle entangled states. (Although \cite{zhang2024bell} claims to use only Bell states, it actually entangles those Bell states with additional particles via CNOT gates and should  be regarded as three-particle entanglement.) Consequently, our protocol requires fewer entanglement resources.

In terms of security analysis, many existing protocols discuss unforgeability only in the sense that an adversary attempts to generate a valid signature from scratch \cite{liu2025quantum,zhang2023semi,zhang2024bell,xia2022semi,zhao2019semi}. However, tampering with an already generated valid signature should also fall within the scope of unforgeability.
For example, in protocol \cite{zhang2024bell}, the security analysis does not take into account the possibility of modifying an existing signature, which leaves a potential security loophole. Specifically, an attacker can apply a bit-flip (Pauli-$X$) operation to  quantum sequences $B$ in signature   without being detected.
In contrast, in the unforgeability proof of our SQS protocol, we explicitly consider both scenarios: forging a signature from scratch and tampering with a  generated valid signature. Therefore, our protocol provides stronger security guarantees.

For the above discussion, a more intuitive comparison is provided in Table \ref{table 2}.

\begin{table}[ht]
\centering
\caption{Comparisons of the similar schemes (Here, “–” indicates that the protocol does not employ an additional eavesdropping-detection mechanism.)}\label{table 2}
\begin{tabular}{lcccc}
\toprule
Schemes 
& \makecell{Number of \\ quantum partners}
& \makecell{Secure eavesdropping \\ detection}
& \makecell{Demand for \\ entangled states}
& Tamper-resistant \\
\midrule
\cite{zhao2019semi} & 2& — & \makecell{Three-particle entangled \\ state}& No\\
\cite{chen2020offline} & 1 & No & \makecell{Four-particle entangled\\ state}& Yes\\
\cite{xia2022semi} & 2 & No & \makecell{Two-particle entangled\\ state}& No\\
\cite{zhang2023semi} & 2 & No & \makecell{Three-particle entangled \\state}& No\\
\cite{zhang2024bell} & 1 & No & \makecell{Three-particle entangled\\ state}&No\\ 
Ours      & 1 & Yes & Bell state&Yes \\
\bottomrule
\end{tabular}

\end{table}
\section{Conclusion}

In this paper, we propose a practical semi-quantum signature protocol based on Bell states in which only the signer must possess full quantum capabilities, while both the receiver and the trusted third party have limited quantum capabilities. This design significantly reduces hardware requirements compared with many existing semi-quantum signature schemes and enhances the feasibility of practical implementation.

To strengthen security in semi-quantum environments, we develop an improved eavesdropping-detection mechanism that combines Z- and X-basis decoy states, decoy reordering, and joint error verification by the sender and the receiver. Unlike conventional direct-reflection approaches, our method effectively resists tampering attacks and eliminates previously overlooked vulnerabilities. In particular, by assigning error-rate verification to both parties, the protocol can detect specific tampering operations, such as Pauli-type modifications, that may bypass earlier detection strategies.

For our SQS protocol, we provide a comprehensive analysis covering correctness, private-key security, unforgeability, undeniability, and resistance to entangling-measurement attacks. By showing that, from the perspective of an attacker, the transmitted quantum sequences are tensor products of maximally mixed states, we prove that the signature achieves information-theoretic security under chosen-plaintext attacks, thereby ensuring the security of the private key. Moreover, in contrast to many existing works that consider only forging signatures from scratch, we explicitly analyze tampering with already generated valid signatures and demonstrate that the proposed protocol remains secure in both scenarios.

Overall, the proposed scheme provides a practical and secure solution for semi-quantum signature scenarios. By combining reduced quantum requirements with enhanced resistance to tampering, it offers a promising direction for future semi-quantum cryptographic applications and contributes to the development of more practical quantum-secure authentication systems.

  \section*{Acknowledgment}

This work was supported by the National Key Research and Development Program of China
under Grant No.2021YFA1000600.

\bibliographystyle{elsarticle-num}
\bibliography{sample}
\biboptions{numbers,sort&compress}

\end{document}